\documentclass[12pt]{iopart}

\usepackage{cite}
\usepackage{hyperref}
\hypersetup{hypertex=true,
colorlinks  =true,
linkcolor           =blue,
anchorcolor         =blue,
citecolor           =blue}
\usepackage{indentfirst}
\expandafter\let\csname equation*\endcsname\relax
\expandafter\let\csname endequation*\endcsname\relax
\usepackage{amsmath}
\usepackage{float}
\usepackage{graphicx}% Include figure files
\usepackage{dcolumn}% Align table columns on decimal point
\usepackage{bm}% bold math
\usepackage{appendix}
\usepackage{verbatim}

\newcommand{\la}{\langle}
\newcommand{\ra}{\rangle}
\newcommand{\sket}[1]{{\ensuremath{\lvert#1\rangle}}}
\newcommand{\lket}[1]{{\ensuremath{\left\lvert#1\right\rangle}}}
\newcommand{\ket}[1]{\if@display\lket{#1}\else\sket{#1}\fi}

\newcommand{\sbra}[1]{{\ensuremath{\langle#1\rvert}}}
\newcommand{\lbra}[1]{{\ensuremath{\left\langle#1\right\rvert}}}
\newcommand{\bra}[1]{\if@display\lbra{#1}\else\sbra{#1}\fi}
\newcommand{\nn}{\nonumber}
\newcommand{\sketbra}[2]{{\ensuremath{\lvert #1\rangle\!\langle #2\rvert}}}
\newcommand{\lketbra}[2]{{\ensuremath{\left\lvert #1\right\rangle\!\!\left\langle #2\right\rvert}}}
\newcommand{\ketbra}[2]{\if@display\lketbra{#1}{#2}\else\sketbra{#1}{#2}\fi}
\usepackage[utf8]{inputenc}
\usepackage[T1]{fontenc}
\usepackage{booktabs}
\usepackage{array, caption, threeparttable}
\usepackage[font=small,labelfont=bf,labelsep=none]{caption}
\makeatletter
\renewcommand{\fnum@figure}{Figure \thefigure.\@gobble}
\makeatother
\usepackage{amsthm}

\usepackage[mathlines]{lineno}% Enable 
\newcommand{\revised}[1]{\textcolor{black}{#1}}

\newtheorem{theorem}{Theorem}
\begin{document}

\title[Quantifying the intrinsic randomness in  sequential measurements]{Quantifying the intrinsic randomness in  sequential measurements}

\author{Xinjian Liu$^{1,2}$, Yukun Wang$^{1,2}$\footnote{wykun06@gmail.com}, Yunguang Han$^{3}$ and Xia Wu$^{4}$}

\address{$^{1}$ Beijing Key Laboratory of Petroleum Data Mining, China University of Petroleum, Beijing 102249, China}
\address{$^{2}$ State Key Laboratory of Cryptology, P.O. Box 5159, Beijing, 100878, China}
\address{$^{3}$ College of Computer Science and Technology, Nanjing University of Aeronautics and Astronautics, Nanjing 211106, China}
\address{$^{4}$ School of Information, Central University of Finance and Economics, Beijing 100081, China}
\ead{wykun06@gmail.com}
\vspace{10pt}

\begin{abstract}
In the standard Bell scenario, when making a local projective measurement on each system component, the amount of randomness generated is restricted. However, this limitation can be surpassed through the implementation of sequential measurements. Nonetheless, a rigorous definition of random numbers in the context of sequential measurements is yet to be established, except for the lower quantification in device-independent scenarios. In this paper, we define quantum intrinsic randomness in sequential measurements and quantify the randomness in the Collins-Gisin-Linden-Massar-Popescu (CGLMP) inequality sequential scenario. Initially, we investigate the quantum intrinsic randomness of the mixed states under sequential projective measurements and the intrinsic randomness of the sequential {positive-operator-valued measure (POVM)} under pure states. Naturally, we rigorously define quantum intrinsic randomness under sequential {POVM} for arbitrary quantum states.
Furthermore, we apply our method to one-Alice and two-Bobs sequential measurement scenarios, and quantify the quantum intrinsic randomness of the maximally entangled state and maximally violated state by giving an extremal decomposition. Finally, using the sequential \revised{Navascues-Pironio-Acin (NPA) }hierarchy in the device-independent scenario, we derive lower bounds on the quantum intrinsic randomness of the maximally entangled state and maximally violated state.
\end{abstract}

% \vspace{2pc}
% \noindent{\it Keywords}: Quantum intrinsic randomness, Quantum random number, Sequential measurement
%\maketitle

\section{Introduction}
Random numbers are essential in information technology, especially information security \cite{stamp2011information}. Many cryptographic protocols \cite{schneier2007applied,abadi1996prudent} require random numbers to prevent attackers from predicting the outcomes of security-related computations to ensure data security. 
\par
In general, there are three main types of random number generators, pseudo-random number generators, classic physical random number generators, and quantum random number generators. However, pseudo-random number generators and classic physical random number generators do not generate truly random numbers, as their randomness cannot be theoretically proven \cite{knuth1981seminumerical}. Quantum random number generators (QRNG) \cite{mayers1998quantum, pironio2010random, lunghi2015self, ma2016quantum, liu2018device} is a device that generates random numbers according to the uncertainty principle of quantum mechanics, ensuring that the generated numbers are truly random.

In practice, quantum devices encounter challenges such as noise and potential third-party interference, necessitating the authentication of the generated random numbers.
Therefore, device-independent protocols \cite{pironio2010random,arnon2018practical,primaatmaja2023security,vazirani2019fully} for random number generation have been proposed. These generators have validation properties that allow them to go through cleverly designed tests, such as \revised{Clauser-Horne-Shimony-Holt (CHSH)} non-locality, to verify that the generated random numbers are truly random and unpredictable and rule out any possible potential attacks or vulnerabilities. Therefore, device-independent quantum true random number generators exhibit higher reliability in terms of security and trustworthiness. However, the tests that device-independent protocols are based on usually result in a relatively low generation rate of verifiable random numbers. 
To enhance the generation rate of verifiable random numbers, researchers have conducted extensive studies and investigations. 
Nonprojective measurements, more specifically positive-operator-valued measures (POVMs)  can generate more randomness by having more outcome possibilities than the dimension of the quantum system they operate on.
In \cite{woodhead2020maximal,acin2016optimal}, the authors proposed a self-testing method based on the nonprojective POVMs, which enables the generation of the optimal possible random numbers consistent with the system's dimensionality.  
In \cite{curchod2017unbounded}, the authors assert that by performing nonprojective measurements sequentially, namely weak measurements \cite{mitchison2007sequential,lundeen2012procedure,gallego2014nonlocality}, on an arbitrarily weakly entangled system, nonlocality can be shared between the sequential pairs of observers. In principle, this sharing of nonlocality between sequential pairs can yield an infinite amount of randomness. 

Subsequently, non-locality sharing in sequential measurements under different entanglement resources and non-locality inequalities have been studied \cite{zhu2022einstein,cheng2021limitations,ren2022nonlocality,zhang2021sharing,pandit2022recycled,steffinlongo2022projective,silva2015multiple,brown2020arbitrarily}, and theoretically, they all can achieve unlimited random number generation. {The measurements in sequential scenarios usually involve POVMs}. However, unlike projective measurements,
there are additional and possibly hidden degrees of freedom in the apparatus for {POVM}.
How to quantify the intrinsic randomness of the
outcomes from {POVM} is an important
and hard problem, given a set of POVM elements
may have an infinite number of ways
to construct the detection instrument \cite{biggerstaff2009cluster,duvsek2002quantum}. This hidden information makes it very challenging to characterize
the amount of information leaked to Eve. \cite{dai2023intrinsic,senno2022quantifying} addressed this problem in non-sequential scenarios. In 
 \cite{dai2023intrinsic}, the intrinsic
randomness for general states under POVM is characterized by minimizing all possible extensions by Naimark extension \cite{bischof2019resource}. And in 
 \cite{senno2022quantifying}, the intrinsic randomness is quantified by introducing an eavesdropper Eve. The quantum intrinsic randomness under  POVM is then obtained according to different degrees of correlation between Eve and quantum systems. However, a rigorous definition of random numbers in the context of sequential measurements is yet to be established, except for the guessing problem quantification in device-independent scenario \cite{bowles2020bounding}. It is also highly meaningful to know how much randomness is in sequential POVMs, particularly in cases where the measurement operations are known but the specific implementation details are unknown.

In sequential scenarios,  the detection instrument decomposition of the POVMs may 
give rise to potential correlations between the sequential measurements, thus the inter-round correlations should be eliminated when characterizing the randomness in the sequential measurements. In this paper, we provide a rigorous definition and quantification of verifiable random numbers generated under different levels of device trustworthiness, encompassing both trusted and untrusted sources  (prepared states and the operated measurements). The organization of the paper is as follows. In section \ref{sectionII}, \revised{we provide two theorems that define the intrinsic randomness of quantum measurement under sequential projective measurement with a shared arbitrary state and sequential POVM with a shared pure state, respectively. Based on the above theorems, we obtain the definition of intrinsic randomness under the sequential POVM with the shared arbitrary state.} In
section \ref{sectionIII}, we apply our method to the Alice, two Bobs sequential scenario and quantify the quantum intrinsic randomness for the maximally entangled state and maximally violated state under \revised{CGLMP}  inequalities. Initially, we examine the quantum intrinsic randomness in the source-trusted case. Subsequently, we obtain bounds on the quantum intrinsic randomness using the sequential NPA method in the device-independent scenario.

\section{Intrinsic randomness in sequential measurements}\label{sectionII}
\subsection{{Preparation with noise}}
\begin{figure}[H]
\centering
\includegraphics[width=0.7\linewidth]{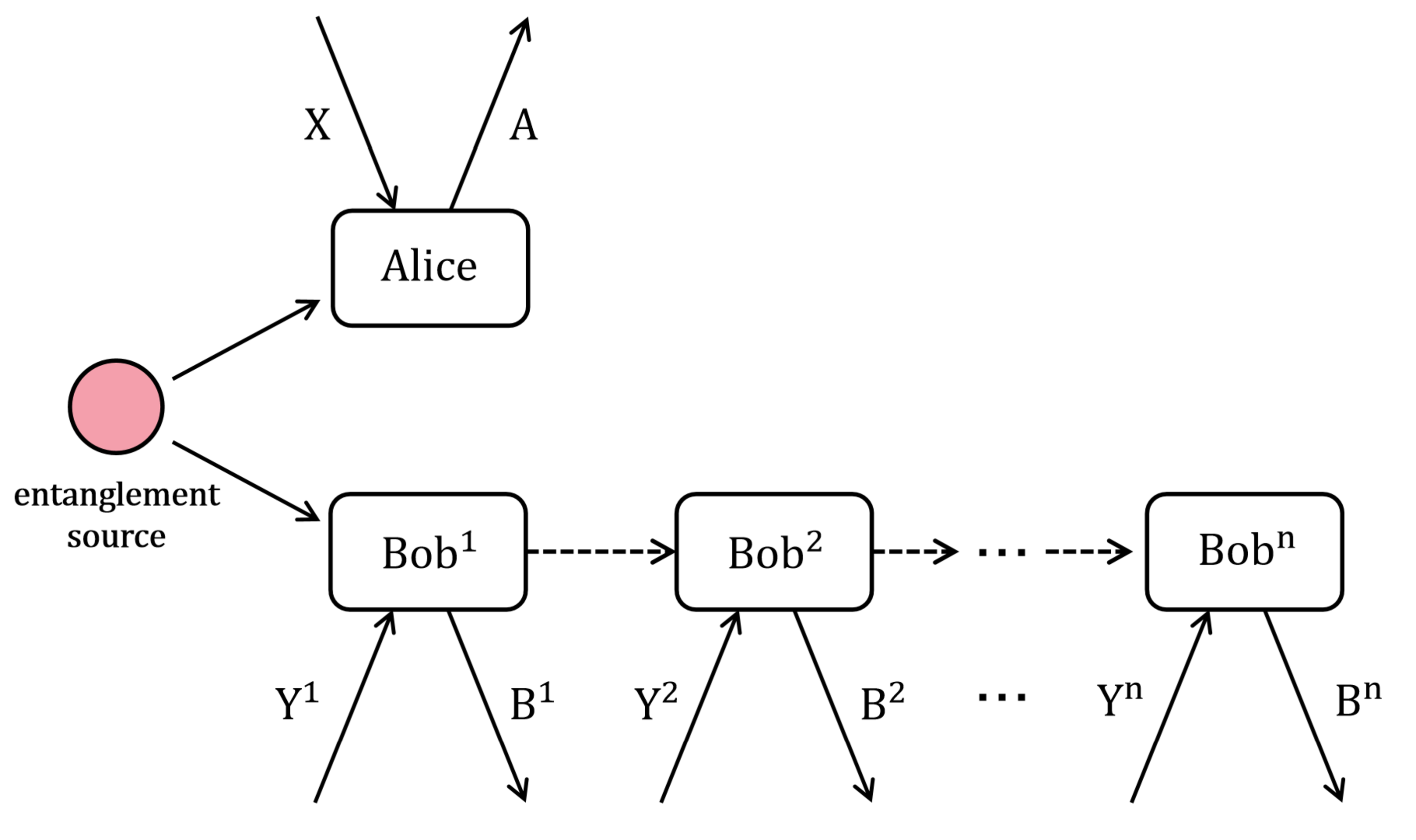} 
\caption{Sequential measurement scenario.}\label{sequential}
\end{figure}

To provide a comprehensive introduction to our work, it is essential to review the concept of sequential measurement. The considered nonlocality sharing scenario is with entangled qubits, where a single
observer, namely Alice, has access to one of the particles of the entangled pair, and a group of observers, Bob$^{i}$ $({i}\in\{1,...,n\})$, has access to the
second particle. Each Bob$^{i}$ acts
independently, performing a local measurement on the particle before passing it on to the next member of the group, see figure \ref{sequential}.
 
We denote the inputs of Alice and Bob$^{i}$ as X and Y$^{i}$, respectively, and their outputs as A and B$^{i}$. 

 Bob's acting independent restriction in the requirement is that each Bob in the sequence can only send a single qubit (his post-measurement state) to the next. In particular, the classical information about measurement choices and outcomes of each Bob is not shared. %It is in this sense that the Bobs act independently of one another. 

Quantum randomness, namely intrinsic randomness, refers to randomness that excludes all possible classical randomness in observed outcomes. To accurately quantify this intrinsic randomness, it is necessary to consider potential noise that may be present in the whole system, including both the state and measurement processes. {Minimum entropy is commonly utilized to characterize the randomness of a probability distribution,  corresponding to the most conservative way of measuring the unpredictability of a set of outcomes}.

To eliminate the influence of known information such as the choice of measurement operation and noise, quantum intrinsic randomness can be characterized using conditional minimum entropy.  Taking into consideration the potential manipulation or prediction of classical and quantum side information by adversaries, this paper explores the classical and quantum correlations between eavesdropper Eve and the system to derive the classical maximum guess probability $P_{\text{guess}}^{C}$ and the quantum maximum guess probability $P_{\text{guess}}^{Q}$ regarding measurement outcomes. Then the randomness in outcomes is defined by the conditional minimum entropy 
\begin{equation}
H_{\infty}(a\Vec{b}|A\vec{B}E) = -\text{log}_{2}[p_{\text{guess}}].
\end{equation}
In our study, we focus on the sequential measurement model given in figure \ref{sequential}, which has been widely proposed and studied in non-local sharing between sequential parties. Note that, it is easily extendable to a simple scenario where quantum states are prepared and only sent to one side multiple Bobs for measurement to obtain randomness, without involving Alice. For instance, in the simplest model, a $\ket{0}$ state is prepared, and multiple Bob sequentially performs measurements using the $X$ and $Z$ bases. In principle, when the devices are trusted (i.e., they indeed prepare the claimed states and perform the specified measurements) and the sequential parties are infinite, an unbounded amount of randomness can be generated. However, in practice, one must also consider the removal of noise in measurements and quantum states to characterize true quantum randomness. Furthermore, if assuming that the devices come from an untrusted third party, then Eve can disturb the devices significantly. 
Therefore, it becomes necessary to consider device-independent scenarios, where sequential non-locality sharing needs to be observed as a measure of quantumness to ensure and quantify randomness. To achieve non-locality sharing in this scenario, each Bob needs to act independently. 
Our definition of randomness is not limited to Bob's independence, which can be extended to scenarios where classical signals are allowed, such as the transmission of measurement basis choices. 

We initially examine the ideal case of the measurements, namely the {measurements without noise} case, to quantify the quantum intrinsic randomness in the outcomes of the sequential measurements. In this case, only the prepared states may have noise, thus we consider the initial state to be mixed. Alice and Bob$^{1}$ sharing a mixed state, with each Bob choosing to perform a projective measurement, represents the ideal scenario that we consider.
When representing the state of an entanglement system $S$ with a mixed state $\rho_s$ that is compatible with an ensemble $\{p(\lambda), |\phi_\lambda\ra\}$. 
Eve may generate classical correlations with quantum systems through random variables $\Lambda=\{\lambda\}$. Specifically, Eve can sample a large number of values of the random variable $\Lambda$ and use this information to predict the outcome of the measurement on the system better than the honest user Alice. Given that a mixed state may consist of many ensembles of pure states, we must consider all possible forms of these ensembles to characterize the quantum intrinsic randomness present in them. This rationale leads to the definition of Eve's classical guessing probability as,
\revised{
\begin{equation}
\begin{split}
    p&^C_{\text{guess}}(\vec{b}|\vec{y},\rho_s,\{\Pi_{y_i}^{b_i}\}_i,E) = \max_{p(\lambda),|\phi_{\lambda}\ra}\sum_{\lambda}p(\lambda)\max_{\vec{b}} \la \phi_{\lambda}| \Pi_{y_1}^{b_1}...\Pi_{y_n}^{b_n}\Pi_{y_{n-1}}^{b_{n-1}}...\Pi_{y_1}^{b_1}|\phi_{\lambda}\ra,\\
\end{split}
\end{equation}}
which maximizes all the pure ensembles of the mixed state, thus characterizing Eve's maximum guessing ability. A larger 
 $p^C_{\text{guess}}$ indicates a stronger guessing ability for Eve, implying her capacity to obtain the randomness of measurement outcomes. The intrinsic randomness in the outcomes of the sequential measurements then is quantified by the conditional minimum entropy of Eve's guessing probability. For Eve's classical guessing probability, it is
 \revised{
\begin{equation}
    H_{\infty}(\vec{b}|\vec{y},\rho_s,\{\Pi_{y_i}^{b_i}\}_i,E)=-\text{log}_2[p^C_{\text{guess}}(\vec{b}|\vec{y},\rho_s,\{\Pi_{y_i}^{b_i}\}_i,E) ],
\end{equation}}
bounding the amount of randomness generated in the worst scenario.
 
Moreover, through the establishment of quantum correlations with the prepared system by purifying the prepared state to $|\psi_{SE}\ra$, Eve could have the potential to acquire quantum side information. By this correlation with the Main system,  Eve could have the state \revised{$\{\rho^{\vec{b}}_{\vec{y}}\}_E = Tr_S[(\Pi^{\vec{b}}_{\vec{y}} \otimes I_E)\ketbra{\psi_{SE}}{\psi_{SE}}]/p(\vec{b},\vec{y})$} after the measurement of \revised{$\Pi^{\vec{b}}_{\vec{y}}$ }on Main system. Usually, the state $\{\rho^{\vec{b}}_{\vec{y}}\}_E$ may not be diagonal on the same basis, then Eve could hold quantum side information more than classical variable $\lambda$. This observation suggests  Eve possesses the capability to manipulate or extract information about $\vec{b}$ from her side of the system. When the mixed state $\rho_{s}$ is being measured, Eve can acquire information about the post-measurement state and subsequently choose measurement operator \revised{$\Pi_E^{\vec{b}}$} that optimally predicts the measurement outcomes of the mixed state.
For the sequential projective measurements \revised{$\{\Pi_{y_i}^{b_i}\}_i$}, the quantum guessing probability of Eve can be mathematically expressed as follows:
\revised{
\begin{equation}\label{Q_guess_noisestate}
\begin{split}
    p^Q_{\text{guess}}(\vec{b}|\vec{y},\rho_s,\{\Pi_{y_i}^{b_i}\}_i,E) =\max_{\{\Pi_E^{\vec{b}}\}_{\vec{b}}}\sum_{\vec{b}} \la \psi_{SE}| \Pi_{y_1}^{b_1}...\Pi_{y_n}^{b_n}I_S\otimes \Pi_E^{\vec{b}}\Pi_{y_n}^{b_n}...\Pi_{y_1}^{b_1}|\psi_{SE}\ra,\\
\end{split}
\end{equation}}
where $|\psi_{SE}\ra$ is any fixed purification of $\rho_s$. Eve
optimizes over measurements \revised{$\Pi_E^{\vec{b}}$} trying to maximize the guessing probability. Once Eve sent the prepared state to the main system $S$, she could not access the main system anymore except to operate the system on her site. 

With the two definitions of the guessing probability of Eve in both classical and quantum correlations, we have the following theorem. 
\begin{theorem}\label{Theorem.1}

For every state and sequential projective measurement
\revised{
    $$p^Q_{\text{guess}}(\vec{b}|\vec{y},\rho_s,\{\Pi_{y_i}^{b_i}\}_i,E) =p^C_{\text{guess}}(\vec{b}|\vec{y},\rho_s,\{\Pi_{y_i}^{b_i}\}_i,E).$$
And the amount of randomness 
    $$H_{\infty}(\vec{b}|\vec{y},\rho_s,\{\Pi_{y_i}^{b_i}\}_i,E)=-\text{log}_2[p^C_{\text{guess}}(\vec{b}|\vec{y},\rho_s,\{\Pi_{y_i}^{b_i}\}_i,E) ] = -\text{log}_2[p^Q_{\text{guess}}(\vec{b}|\vec{y},\rho_s,\{\Pi_{y_i}^{b_i}\}_i,E) ].$$
}
    
\end{theorem}
The proof is
given in the appendix.
It implies that, under the sequence of projective measurements, when considering only the correlations of Eve on the prepared state, both her quantum-capable and classical-capable abilities result in an equal probability of successfully obtaining the measurement outcomes, \revised{and thus generate the same amount of randomness.}

\subsection{Sequential {measurements with noise} }

In this section, we examine the influence of sequential POVM on characterizing random numbers. Before presenting our contributions, it is important to review how sequential POVM impacts the characterization of randomness.
When give a quantum state and a POVM set%$\{F_{0}, F_{1}\}$
, it is necessary to eliminate classical randomness from it.  The POVM, similar to a mixed state, possesses unitary degrees of freedom, resulting in various decompositions. To remove the classical randomness, all possible forms of POVM decomposition must be taken into account.
In the context of sequential {POVM}, the analysis of randomness requires considering not only different forms of POVM decomposition but also the intercorrelations between sequential measurements.

%In the context of this scenario, it is important to note that two-dimensional POVM (Positive Operator-Valued Measure) can be seen as a combination of projection measurements and noise. This noise component may arise due to the presence of Eve, who is an eavesdropper or an external entity.

Consider the most commonly concerned case, CHSH non-locality sharing. As illustrated in 
 \cite{brown2020arbitrarily}, the setup is one-sided sequential measurements where one of the parties has two independent observers. Alice and Bob$^{1}$ share the pure two-qubit state $|\phi\rangle=\frac{1}{\sqrt{2}}(|00\rangle + |11\rangle)$ and  the  considered measurements strategy for non-locality sharing are as follows, \revised{
$\{  A_{0|0} = \frac{I+\cos(\theta)\sigma_{z}+\sin(\theta)\sigma_{x}}{2},
    A_{0|1} = \frac{I+\cos(\theta)\sigma_{z}-\sin(\theta)\sigma_{x}}{2}.\}$
and $\{
    B_{0|0}^{1} = \frac{I+\sigma_{x}}{2},
    B_{0|1}^{1} = \frac{I+\varepsilon_0 \sigma_{z}}{2};
    B_{0|0}^{2} = \frac{I+\sigma_{x}}{2},
    B_{0|1}^{2} = \frac{I+ \varepsilon_1\sigma_{z}}{2}
        \label{bob2}\}$. The subscript $\{b|y\}$ of $B$ represents the outcomes and the measurement chosen by Bob, while the superscript of $B$  represents the sequence number of Bob.}
We consider the scenario where there are two sequential on Bob's side and focus on the measurements of B$_{0|1}^{1}$ and B$_{0|1}^{2}$.  
The resulting post-measurement state is determined by the decomposition of Bob$^{1}$.
After Bob$^{1}$ has been measured, it passes the post-measurement state to Bob$^{2}$. Then Bob$^{2}$ can independently select one of the two measurements as defined to measure.
To lower bound the randomness, we should allow Eve to implement the POVMs arbitrarily. Assume that Eve characterizes the observed probability distribution through convex decomposition,
and has the following decomposition,
\begin{equation}
\text{POVM}_{\{B_{i|1}^{1}\}}=\varepsilon_{i}P_0+{(1-\varepsilon_{i})}P_1,
\end{equation}
where
$P_0=\{\ketbra{0}{0},\ketbra{1}{1}\}$, and $P_1=\{\ketbra{1}{1},\ketbra{0}{0}\}$, which have been proved to be standard decomposition form for qubit system \cite{cao2015loss}.
 
 Despite the independence and irrelevance of Bob's actions, Eve may possess different  $\varepsilon_{i}$, which can affect the subsequent choices based on the previous round's selection, rendering Bob unable to detect this correlation. 
For instance, when \revised{
$\varepsilon_0=\varepsilon_1$}, Eve selects the projective measurement $P_0$ for both 
Bob$^1$ and Bob$^2$ with a probability of \revised{$\varepsilon_0$}. Conversely, with a probability of 
\revised{$1-\varepsilon_0$}, Eve opts for the projective measurement 
$P_1$ for both Bob$^1$ and Bob$^2$. In this strategy, once Eve correctly guesses  $b_1$, she obtains the knowledge of the outcome of $b_2$. 
 Thus, the observed probabilities in this case, which are,
 \revised{
\begin{equation}
\begin{split}    p(b_1,b_2|B_1^1,B_1^2)&=\varepsilon_0 \varepsilon_1  \la \phi| {P_{0}^{b_1}}{P_{0}^{b_2}}{P_{0}^{b_1}}|\phi\ra.\\
    &+\varepsilon_0 (1-\varepsilon_1 ) \la \phi| {P_{0}^{b_1}}{P_{1}^{b_2}}{P_{0}^{b_1}}|\phi\ra.\\
    &+(1-\varepsilon_0 )\varepsilon_1  \la \phi| {P_{1}^{b_1}}{P_{0}^{b_2}}{P_{1}^{b_1}}|\phi\ra.\\
    &+(1-\varepsilon_0) (1-\varepsilon_1 ) \la \phi| {P_{1}^{b_1}}{P_{1}^{b_2}}{P_{1}^{b_1}}|\phi\ra,\\
    \end{split}
\end{equation}}
  cannot be directly used to define the randomness by their min-entropy. The influence of the classical variables of $\varepsilon$ needs to be eliminated.  In the following section, we will focus on addressing the methods to resolve this issue, and then we will know the randomness should be quantified by the guessing probability of
\begin{equation}
    p_{\text{guess}}(b_1,b_2|B_1^1,B_1^2,E)=\text{max}_{{b_1}}  \la \phi| {P_{0}^{b_1}}|\phi\ra,
\end{equation}
which is the correctly guessing probability of Eve about Bob's outcomes, given the measurements 
$B_1^1, B_1^2$, and the knowledge she possesses (represented by the variable E).
Before exploring the most general scenario, we examine sequential {POVM} applied to a pure-state system. To maximize the preservation of entanglement in the post-measurement state, we utilize unsharp measurement for the quantum system.
We examine a scenario wherein sequential {POVMs}, represented by POVM 
$\{M_S^{b_i}\}_{b_i}$, are performed on the main system 
$S$ that is initially in a pure state  $|\psi_S\ra$.
Given that the set of POVMs is, like the set of quantum states, convex, we can proceed via analogy with the case of a mixed state. Assuming Eve possesses the ability to sample a random variable \revised{$\omega$} such that \revised{$M_S^{b_i} = \sum_{\omega_i}p(\omega_i)M_S^{b_i,\omega_i}$ with $\{M_S^{b_i,\omega_i}\}_{b_i}$},which are \revised{projection-valued measures (PVMs)}, for all \revised{$\omega_i$}.  
With her knowledge of \revised{$\omega_i$}, her optimal prediction for the outcome of the measurement on $S$ is  is represented by $p_{\text{opt}}^{\text{c}}$, which can be calculated as \revised{$p_{\text{opt}}^{\text{c}}$ = $\max_{\vec{b}}\la \psi_S|\sqrt{M_S^{b_1,\omega_1}}^\dagger...M_S^{b_n,\omega_n}...\sqrt{M_S^{b_1,\omega_1}}|\psi_S\ra$.}

We can consequently establish 
 Eve's maximum classical correctly guessing probability, by optimizing all possible convex decomposition of the POVMs. The maximum classical guessing probability of Eve is 
% \begin{equation*}
% \begin{split}
%     p&^C_{\text{guess}}(\vec{b}|\vec{y},|\psi_S\ra,\{M_{S}^{b_i}\}_i,E) \\
%     &= \max_{p_{\lambda_i},\{M_S^{b_i,\lambda_i}\}_{b_i}}\sum_{\lambda_1}p_{\lambda_1}...\sum_{\lambda_n}p_{\lambda_n}\\
%     &\cdot\max_{\vec{b}} \la \psi_S| \sqrt{M_{y_1}^{b_1,\lambda_1}}^\dagger...M_{y_n}^{b_n,\lambda_n}...\sqrt{M_{y_1}^{b_1,\lambda_1}}|\psi_S\ra\\
% \end{split}\label{pguess_C_noiseM}
% \end{equation*}
\revised{
\begin{equation}
\begin{split}
    p^C_{\text{guess}}&(\vec{b}|\vec{y},|\psi_S\ra,\{M_{S}^{b_i}\}_i,E) = \max_{p(\omega_i),\{M_S^{b_i,\omega_i}\}_{b_i}}\sum_{\omega_1,...,\omega_n}p_{\omega_1,...,\omega_n} \max_{\vec{b}} \la \psi_S| \sqrt{M_{y_1}^{b_1,\omega_1}}^\dagger...M_{y_n}^{b_n,\omega_n}...\sqrt{M_{y_1}^{b_1,\omega_1}}|\psi_S\ra.\\
\end{split}\label{pguess_C_noiseM}
\end{equation}}
% where 
% \begin{align}
% &\sum_{\lambda_i,i\neq j}p_{\lambda_1,...,\lambda_n}=p_{\lambda_j},\\  \nn &p_{\text{opt}}^{\text{c}} = \max_{\vec{b}} \la \psi_S| \sqrt{M_{y_1}^{b_1,\lambda_1}}^\dagger...M_{y_n}^{b_n,\lambda_n}...\sqrt{M_{y_1}^{b_1,\lambda_1}}|\psi_S\ra.
% \end{align}

\begin{figure}[H]
\centerline{\includegraphics[width=0.7\linewidth]{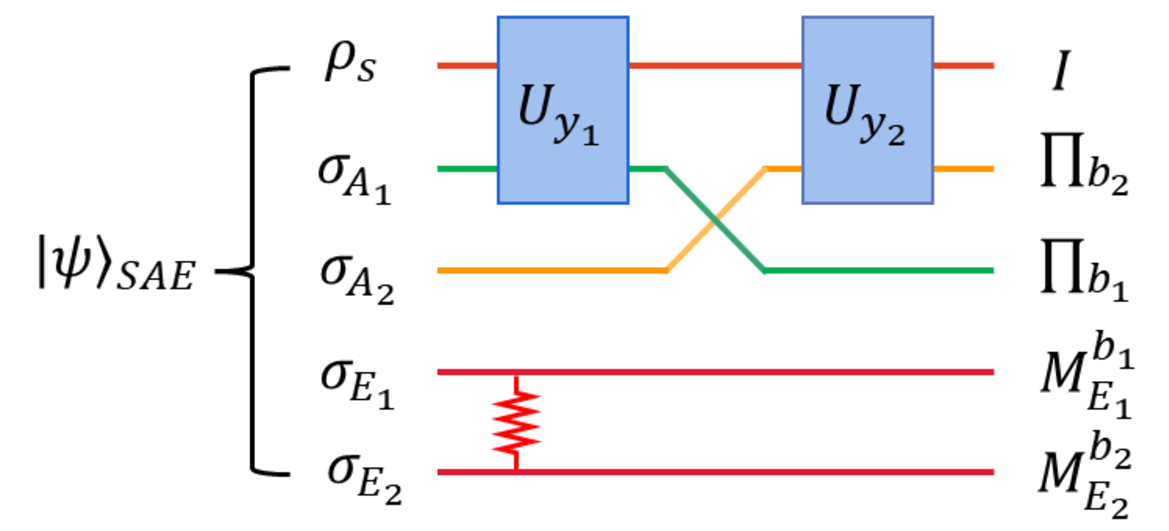}}
\caption{The adversary scenario for a generalized Naimark
extension. The projective measurement $\Pi_{y_i}^{b_i}$
 which has the formula $\Pi_{y_i}^{b_i} = U_{y_i}^\dagger (I\otimes \Pi_{b_i})
U_{y_i}$ gives a projective extension of $M_{S}^{b_{i}}$. By tracing out the ancillary system and Eve,  $M_{S}^{b_{i}}$ is recovered.  }\label{figsequentialM}
\end{figure}
Subsequently, we broaden the analysis to the definition of quantum guessing probabilities about sequential measurements.  Quantum guessing probabilities were initially introduced by Frauchig et al. in their work \cite{frauchiger2013true}, wherein they examined general POVMs (non-sequential scenarios) using the Naimark extension. We extend their approach and apply it to the definition of quantum guessing probabilities under sequential measurements. According to the Naimark extension, a general measurement of the main system can be regarded as a project measurement of the composite system consisting of the main system and an auxiliary system A as shown in figure \ref{figsequentialM}.
$\{\Pi_{y_i}^{b_i}\}_{b_{i}}$ which acts on joint system $SA_i$, is a Naimark extension of $M_{S}^{b_{i}}$, and the correlations with Eve are represented by a mixed state $\sigma_{A}$ on the auxiliary system A, for which she possesses a purification 
$|\phi_{A_iE_i}\ra$. The purification process enables Eve to establish an entanglement relationship with both the system S and the ancilla A. This allows Eve to gain access to the quantum side information of the main system. Eve then performs optimizations on measurements conducted on her subsystem.
We could introduce one Eve for each individual within the sequence, denoted as $M_{E_i}^{b_i} $, to guess the outcomes of Bob$^i$'s outcomes. However, notice that we allow Eves to exhibit interactions among themselves. For such scenarios, instead of introducing individual measurement, we introduce a joint measurement $M_{\vec{E}}^{\vec{b}} $ that offers  $2^n$ possible outcomes to characterize Eve's capacity. This may give Eve more information than individual measurements.
Aiming to maximize the correlation with the measurement outcomes of the user, the maximal quantum guessing probability for Eve can be determined as follows:

\begin{equation}
\begin{split}\label{Q_guess_noiseM}
            &p^Q_{\text{guess}}(\vec{b}|\vec{y},|\psi_S\ra,\{M_{S}^{b_i}\}_i,E) 
    = \max_{\{\Pi_{y_i}^{b_i}\}_{b_i},|\phi_{A_iE_i}\ra,\{M_{E_i}^{b_i}\}_{b_i}}\sum_{\vec{b}} (\prod_{i} \la \psi_S| \la \phi_{A_iE_i}|(\Pi_{\vec{y}}^{\vec{b}})^\dagger\otimes M_{E_i}^{b_i} \Pi_{\vec{y}}^{\vec{b}} 
 | \phi_{A_iE_i}\ra|\psi_S\ra,\\
% &\text{where}\\
%    &p_{\text{opt}}^{\text{q}} = \sum_{\vec{b}} (\prod_{i} \la \psi_S| \la \phi_{A_iE_i}|(\Pi_{\vec{y}}^{\vec{b}})^\dagger\otimes M_{E_i}^{b_i} \Pi_{\vec{y}}^{\vec{b}} 
%  | \phi_{A_iE_i}\ra|\psi_S\ra. \\
&\text{ subject to}\\
&\Pi_{\vec{y}}^{\vec{b}}={\Pi_{y_n}^{b_n}}...\Pi_{y_1}^{b_1}\\
&\tr_{A_i}[\Pi^{b_i}
_{SA_i}(I_S \otimes  \tr_{E_i}[| \phi_{A_iE_i}\ra\la \phi_{A_iE_i}|])] = M_S^{b_i}, \;\;
\forall b_i.\\
&\tr(\Pi_{y_i}^{b_i}\otimes I \ketbra{\Pi_{y_{i-1}}^{b_{i-1}},...,\Pi_{y_{1}}^{b_{1}}\psi_{S}}{\Pi_{y_{i-1}}^{b_{i-1}},...,\Pi_{y_{1}}^{b_{1}}\psi_{S}})=\tr(M_{y_i}^{b_i}\rho_{S}^{\text{post}}),\\
% &\rho_{S}^{\text{post}}=\sum_{\lambda_1,...,\lambda_n}p(\lambda_{i-1},...,\lambda_1)\sqrt{M_{y_{i-1}}^{b_{i-1},\lambda_{i-1}}},...,\sqrt{M_{y_{1}}^{b_{1},\lambda_{1}}}\rho_S\\
% & \;\;\;\;\;\;\cdot\sqrt{M_{y_{1}}^{b_{1},\lambda_{1}}},...,\sqrt{M_{y_{i-1}}^{b_{i-1},\lambda_{i-1}}}
& \text{with}\;\;\rho_{s}^{\text{post}} = |\phi\rangle \langle \phi|_{\text{post}},\text{where}\\
&|\phi\rangle_{\text{post}} = \sum_{\lambda_{i},...,\lambda_{i-1}} \sqrt{M_{y_{n-1}}^{b_{n-1},\lambda_{i-1}}},...,\sqrt{M_{y_{1}}^{b_{1},\lambda_{1}}}|\psi_S\rangle,\\
\end{split}
\end{equation}
where $\Pi_{\vec{y}}^{\vec{b}}={\Pi_{y_n}^{b_n}}...\Pi_{y_1}^{b_1}$ represents the sequential projective measurements performed on the received state. The state $|\phi\rangle_{\text{post}} $ denotes the post-measurement state of the first $i$ participant in Bob's site. 

With the classical guessing probability and the quantum guessing probability of Eve for sequential {POVM} in the pure state case, \revised{their capabilities are equivalent, and the use of conditional minimum entropy to describe the intrinsic randomness of quantum is equivalent.} we have the following theorem. 
\begin{theorem}\label{Theorem.2}
For every pure state $|\psi_{s}\rangle $ and every sequential POVM.

   $$p^C_{\text{guess}}(\vec{b}|\vec{y},|\psi_S\ra,\{M_{S}^{b_i}\}_i,E)=p^Q_{\text{guess}}(\vec{b}|\vec{y},|\psi_S\ra,\{M_{S}^{b_i}\}_i,E).$$
\revised{
And the amount of randomness 
    $$H_{\infty}(\vec{b}|\vec{y},|\psi_S\ra,\{M_{S}^{b_i}\}_i,E)=-\text{log}_2[p^C_{\text{guess}}(\vec{b}|\vec{y},|\psi_S\ra,\{M_{S}^{b_i}\}_i,E)] = -\text{log}_2[p^Q_{\text{guess}}(\vec{b}|\vec{y},|\psi_S\ra,\{M_{S}^{b_i}\}_i,E)].$$}
\end{theorem}
We give the proof in the Appendix.

\subsection{{Both preparation and measurements with noise}}
After \revised{defining} the randomness in the presence of {noise} in the sequential measurements scenario mentioned above under pure state, we now proceed to the most general setting. We consider a sequential sequence of POVMs $\{M_{y_i}^{b_i}\}_{b_i}$, where $i=1,...,n$, being measured sequentially on the system $S$ in a noisy state $\rho_S$. 
When taking into account classical side information, Eve selects convex decompositions of both the state and the measurement. In this case, her classical guessing probability is given by
\begin{equation}
\begin{split}
    p&^C_{\text{guess}}(\vec{b}|\vec{y},\rho_S,\{M_{y_i}^{b_i}\}_i,E) \\
    &= \max_{p_{\lambda_S,\lambda_1,...,\lambda_n},\{|\psi_{\lambda_S}\ra\}_{\lambda_S}, \{M_S^{b_i,\lambda_i}\}_{b_i}}\sum_{\lambda_S,\lambda_1,...,\lambda_n}p_{\lambda_S,\lambda_1,...,\lambda_n}\max_{\vec{b}} \la \psi_{\lambda_S}| \sqrt{M_{y_1}^{b_1,\lambda_1}}^\dagger...M_{y_n}^{b_n,\lambda_n}...\sqrt{M_{y_1}^{b_1,\lambda_1}}|\psi_{\lambda_S}\ra\\
    &\text{ subject to}\\
&\sum_{\lambda_S,\lambda_1,...,\lambda_n}p_{\lambda_S,\lambda_1,...,\lambda_n}|\psi_{\lambda_S}\ra\la \psi_{\lambda_S}| =\rho_S\\
&\sum_{\lambda_S,\lambda_1,...,\lambda_n}p_{\lambda_S,\lambda_1,...,\lambda_n}\sqrt{M_{y_i}^{b_i,\lambda_i}} = M_{y_i}^{b_i}, \;\;
\forall b_i,y_i.
\end{split}\label{pguess_C_noiseM-P}
\end{equation}
A joint probability $p_{\lambda_S,\lambda_1,...,\lambda_n}$ is introduced to represent that Eve could manipulate the state and sequential measurements with classical side information. 
Furthermore, let Eve have the quantum correlation with the state and the POVM. Then she could introduce a system that purifies the state and holds the purification $|\psi_{SAE}\rangle$ of the noisy state $\rho_S$ and ancillary system A for quantum side information. Consider the Naimark extension of the POVMs at the same time,  Eve's quantum guessing probability is given by

\begin{equation}
\begin{split}\label{Q_guess_noiseM-P}
 p&^Q_{\text{guess}}(\vec{b}|\vec{y},\rho_S,\{M_{S}^{b_i}\}_i,E) \\
    &= \max_{\{\Pi_{SA_i}^{b_i}\}_{b_i},|\psi_{S\vec{A}\vec{E}}\ra,\{M_{E_i}^{b_i}\}_{b_i}}\sum_{\vec{b}} \la \psi_{S\vec{A}\vec{E}}| (\Pi_{\vec{y}}^{\vec{b}})^\dagger\otimes M_{E_1}^{b_1}\otimes...\otimes M_{E_n}^{b_n} \Pi_{\vec{y}}^{\vec{b}} 
\ra |\psi_{S\vec{A}\vec{E}}\ra. \\
&\text{ subject to}\\
&\tr_{\vec{A}\vec{E}}\ketbra{\psi_{S\vec{A}\vec{E}}}{\psi_{S\vec{A}\vec{E}}}=\rho_S\\
&\Pi_{\vec{y}}^{\vec{b}}={\Pi_{y_n}^{b_n}}...\Pi_{y_1}^{b_1}\\
&\tr_{A_i}[\Pi^{b_i}
_{SA_i}(I_S \otimes  \tr_{SE,A/{A_i}}(\ketbra{\psi_{S\vec{A}\vec{E}}}{\psi_{S\vec{A}\vec{E}}}))] = M_{y_i}^{b_i}, \;\;
\forall b_i,y_i.\\
&\tr(\Pi_{y_i}^{b_i}\otimes I \ketbra{\Pi_{y_{i-1}}^{b_{i-1}},...,\Pi_{y_{1}}^{b_{1}}\psi_{S\vec{A}\vec{E}}}{\Pi_{y_{i-1}}^{b_{i-1}},...,\Pi_{y_{1}}^{b_{1}}\psi_{S\vec{A}\vec{E}}})=\tr(M_{y_i}^{b_i}\rho_{S}^{\text{post}}), \text{with}\\
&\rho_{S}^{\text{post}}=\sum_{\lambda_1,...,\lambda_n}p(\lambda_{i-1},...,\lambda_1)\sqrt{M_{y_{i-1}}^{b_{i-1},\lambda_{i-1}}},...,\sqrt{M_{y_{1}}^{b_{1},\lambda_{1}}}\rho_S\sqrt{M_{y_{1}}^{b_{1},\lambda_{1}}},...,\sqrt{M_{y_{i-1}}^{b_{i-1},\lambda_{i-1}}}
\end{split}
\end{equation}
In principle, in general scenario, both the state and measurement are noisy, typically satisfying  $p^Q_{\text{guess}}(\vec{b}|\vec{y},\rho_S,\{M_{S}^{b_i}\}_i, E) \geq  p^C_{\text{guess}}(\vec{b}|\vec{y},\rho_S,\{M_{y_i}^{b_i}\}_i, E)$. This advantage often relies on the entanglement between S and A, where Eve can gain an advantage by measuring her system, obtaining more information about the main system than the classical side information scenario. However, when the optimal value results in the separable post-measurement state with Eve's measurement, the advantage of quantum measurement may no longer exist. The proof is similar to Theorem \ref{Theorem.2} in the appendix, but with a modification to consider mixed states instead of pure states.

\section{Application: The randomness in shared entanglement scenario under CGLMP inequality}\label{sectionIII}
\subsection{\label{sec:level2}CGLMP inequality}
Bell's inequality forms the basis for the study of quantum entanglement, non-locality, and the secure certification of random numbers in the device-independent scenario.
 To deepen our understanding of entanglement and non-locality, various generalized Bell inequalities have been derived. The CGLMP inequality is an example specifically designed for high-dimensional quantum systems. It takes the form:
\revised{
\begin{equation}
 I_{d} = \sum\limits_{k = 0}\limits^ {\lfloor\frac{d}{2}-1\rfloor}(1-\frac{2k}{d-1})[f(k)-f(-k-1)],\label{high-dim}
\end{equation}}
where 
\begin{equation}
\begin{split}
    f(k) &= P(A_{1}  = B_{1} + k ) + P(B_{1}  = A_{2} + k +1 )  + P(A_{2}  = B_{2} + k ) + P(B_{2}  = A_{1} + k ),
\end{split}
\end{equation}
and
\revised{
\begin{equation}
    P(A_{1} = B_{1}+k) = \sum\limits_{j=0}\limits^{d-1}P(A_{1} = j,B_{1} = (j+k)\bmod d),
\end{equation}
where $P(A_{1} = B_{1}+k)$ that the measurements $A_{1}$ and $B_{1}$ have outcomes that differ, modulo $d$, by $k$.}
\par
Alice and Bob are two observers, each equipped with the capability to perform two measurements, each yielding $d$ outcomes. A$_{1}$ and A$_{2}$ represent the measurement settings of Alice, and B$_{1}$ and B$_{2}$ represent the measurement settings of Bob.
In this paper, our primary focus lies on the CGLMP inequality in 3-dimensional quantum systems. Specifically, the CGLMP inequality for qutrits, which correspond to 3-dimensional quantum systems, can be derived from equation (\ref{high-dim}).
\begin{equation}
\begin{split}
    I_{3} &= P(A_{1} = B_{1}) + P(B_{1} = A_{2}+ 1)\\
    &+P(A_{2} = B_{2}) + P(B_{2} = A_{1}) \\
    &-P(A_{1} = B_{1}-1) - P(B_{1} = A_{2})\\
    &-P(A_{2} = B_{2}-1) - P(B_{2} = A_{1} - 1).
\end{split}
\end{equation}
\par
An intriguing fact is that, for dimensions $d\geq 3$, the quantum state that results in the maximum violation of the CGLMP inequality is not the maximally entangled state (MES). Rather, the optimal state is known as the maximum violation state (MVS). 
The form of the maximally entangled state in a three-dimensional quantum system is given by:
\begin{equation}
    |\psi_{\text{MES}}\rangle = \frac{1}{\sqrt{3}}(|00\rangle+|11\rangle+|22\rangle)\label{MES}.
\end{equation}
The maximum violation state is 
\begin{equation}
    |\psi_{\text{MVS}}\rangle = \frac{1}{\sqrt{2+\gamma^{2}}}(|00\rangle+\gamma|11\rangle+|22\rangle),\label{MVS}
\end{equation}
where $\gamma=(\sqrt{11}-\sqrt{3})/2$.

In this context, the
optimal violation of the CGLMP inequality in a 3-dimensional quantum system is achieved by selecting local measurements. These measurements can be described by the \revised{operators} $A_{x}$ and $B_{y}$, along with their associated eigenvectors, where $x\in\{0,1\}$ and $y\in\{0,1\}$. It takes the form:
\revised{
\begin{equation}
    |k\rangle_{A_x} = \frac{1}{\sqrt{3}}\sum\limits_{j=0}\limits^{2}\exp(\frac{2\pi i}{3}j(k+\alpha_{x}))|j\rangle_{A}.\label{Cglmp_a}
\end{equation}
\begin{equation}
    |l\rangle_{B_y} = \frac{1}{\sqrt{3}}\sum\limits_{j=0}\limits^{2}\exp(\frac{2\pi i}{3}j(-l+\beta_{y}))|j\rangle_{B}.\label{Cglmp_b}
\end{equation}
with 
\begin{equation}
    \alpha_{0} = 0, \quad \alpha_{1} = \frac{1}{2}, \quad \beta_{0} = \frac{1}{4} \quad \text{and} \quad  \beta_{1} = -\frac{1}{4}
    \label{cglmp_par}
\end{equation}}

\subsection{{Sharing qutrit non-locality in one-sided sequential measurements}}
The violation of Bell's inequality in quantum systems serves as a witness for quantum non-locality. Quantum non-locality and entanglement are the foundation for the generation of quantum randomness. The relation between randomness and nonlocality is an interesting issue. The researchers came up with a non-local guessing game to characterize the relationship. In this work, we investigate the violation of the CGLMP inequality by the maximum entangled state and the maximum violated state in a 1-Alice and 2-Bobs sequence scenario. The measurement settings at Alice are denoted by $|k\rangle_{A}\langle k|_{A}$ and the  weak measurements on the Bob$^{m}$ side is in the form:
\begin{equation} \label{white-noisy measurement}   
    E_{B^{m}_y}^{l} = \varepsilon_{m}|l\rangle_{B,y}\langle l|_{B,y}+\frac{1-\varepsilon_{m}}{3}I_{3}.
\end{equation}
For each participant Bob$^{m}$, the two measurements have introduced the same noisy parameter $\varepsilon_{m}$.
Alice and Bob$^{1}$ shared the maximally entangled state (as described in equation (\ref{MES})). Then, Alice and Bob$^{1}$ performed local measurements for the CGLMP test, as described in equations (\ref{Cglmp_a}), (\ref{Cglmp_b}), (\ref{cglmp_par})
 and (\ref{white-noisy measurement}). The value of the violated CGLMP inequality is:
\begin{equation}
    I_{3}^{1} = \frac{4}{9}(3+2\sqrt{3})\varepsilon_{1}.
\end{equation}
\par
When the coefficient of weak measurement exceeds 0.69615, the correlation between Alice and Bob$^{1}$ can violate the CGLMP inequality. When $\varepsilon_{1} = 1$,  the optimal violation can be achieved. To maximize the non-locality shared between Alice and Bob$^{2}$, Bob$^{1}$ uses the weak measurement. As a result, when Bob$^{2}$ performed local measurement described in equation (\ref{Cglmp_b}), the quantum expression for Alice and Bob$^{2}$ is obtained as follows :
\begin{equation}
\begin{split}
    I_{3}^{2} &= \frac{1}{81}(56\sqrt{3} - (24+8\sqrt{3}\varepsilon_{1})+ (48+16\sqrt{3})\sqrt{1-\varepsilon_{1}}\sqrt{1+2\varepsilon_{1}}+60).
\end{split}
\end{equation}

Similarly, we also consider the case that the shared quantum state is the maximum violation state. Alice and Bob$^{1}$ shared the maximal violation states (equation (\ref{MVS})), Alice and Bob$^{1}$ performing measurement for the CGLMP test given in equations (\ref{Cglmp_a}), (\ref{Cglmp_b}), (\ref{cglmp_par}), The violation value of the CGLMP inequality between Alice and Bob$^{1}$ can be obtained through calculation.
\begin{equation}
    I_{3}^{1} = (1+\sqrt{\frac{11}{3}})\varepsilon_{1}.
\end{equation}
Alice and Bob$^{2}$ violated CGLMP inequality value is 
\begin{equation}
    I_{3}^{2} = 1.929 + 0.986 \sqrt{(1-\varepsilon_{1})(1+2\varepsilon_{1})} -0.493 \varepsilon_{1}.
\end{equation}
% In MVS, we also get double violation when the noisy is 
% \begin{equation}
%     0.6861<\lambda_{1} < 0.9024
% \end{equation}
\begin{table}
\centering
\setlength{\abovecaptionskip}{10pt}
\setlength{\belowcaptionskip}{10pt}
\caption{The maximum violation values of CGLMP that can be achieved by $\text{Bob}^{1}$ and $\text{Bob}^{2}$ in sequential measurements for the maximum entangled state and maximum violated state. The range of values for the weak measurement of the $\text{Bob}^{1}$ side to achieve maximum shared nonlocality.}\label{bs}
\begin{tabular}{cccc}
\toprule
 & $\text{Bob}^{1}$ & $\text{Bob}^{2}$ & double violations\\
\midrule
MES   & 2.8729     & 2.4086 &  0.696 \textless $\varepsilon_{1}$ \textless 0.904  \\
\midrule
MVS&  2.915  & 2.440  & 0.686 \textless $\varepsilon_{1}$ \textless 0.902 \\
\bottomrule
\end{tabular}

\end{table}

\subsection{Randomness in observed statistics }

In this section, we investigate the generation of random numbers under the CGLMP sequential measurement scenario discussed above, within the noise interval where both $I^1_3$ and $I^2_3$ violations. We will analyze both trustworthy states and measurements, as well as untrustworthy states and measurements.
\subsubsection{quantum randomness in trusted quantum system scenario}
The trustworthiness of the states and measurements refers to the knowledge of their density operator and POVM form, while their specific decomposition forms are unknown. We consider the state to be in a pure form. Based on our definition of randomness, when the state is pure, the quantum guessing probability and classical guessing probability are equal. Therefore, to quantify the randomness, it is necessary to explore all possible decomposition forms of measurements. The average probability with which the verifier
can guess the outcomes observed by Alice and Bob correctly, given his knowledge of the inputs is
thus given by the maximum of

\begin{equation}
G(a,\vec{b}|x,\vec{y},\lambda_{i})= \sum_{i}\lambda_{i}\max_{a,\vec{b},\{P_{y_1}^{b_1}\}_{\lambda_i}}\tr(\rho_{AB}\{P_{y_1}^{b_1}\}_{\lambda_i}) \tr(\{P_{y_1}^{b_1}\}_{\lambda_i}\rho_{AB}\{P_{y_1}^{b_1}\}_{\lambda_i}^\dagger M_{x}^{a}\otimes M_{y_2}^{b_2}). 
\label{global guess probability}
\end{equation} 

 As mentioned, $M_{y_2}^{b_2}$ is a projective measurement, thus in the definition of the max function is only to run over all possible POVM decomposition of $M_{y_1}^{b_1}$. And we denoted $M_{y_1}^{b_1}$to be  $M_{y_1}^{b_1}=\sum_{i}\lambda_i\{P_{y_1}^{b_1}\}_{\lambda_1}$. 
 Typically, in non-locality sharing scenarios, the last measurement in the sequence is often a projective measurement, as it is necessary to destroy the entanglement at the end to obtain maximum non-locality.  Regarding the generation of randomness, performing the projective measurement in the last sequence can also extract additional residual randomness from the final post-measurement state. 
 
In principle, quantifying randomness requires exploring all extreme value decompositions of POVM. However, exploring all extreme value decompositions is challenging, especially for high-dimensional systems, and it is unclear how many extreme value decompositions a POVME  can have.
 To quantify the amount of quantum randomness involved in the sequential CGLMP scenario, we provide a specific extreme POVM decomposition for the POVM defined in equation (\ref{white-noisy measurement}), with the following decomposition:
\begin{equation}
\text{POVM}_{\{E_{B^{m}}^{l}\}}=\varepsilon_{m}P_0+\frac{1-\varepsilon_{m}}{3}P_1+\frac{1-\varepsilon_{m}}{3}P_2+\frac{1-\varepsilon_{m}}{3}P_3 \label{extremal_de},
\end{equation}
where
$P_0=\{\ketbra{0}{0},\ketbra{1}{1},\ketbra{2}{2}\}$,

$P_1=\{\ketbra{\theta_0}{\theta_0},\ketbra{\theta_1}{\theta_1},\ketbra{\theta_2}{\theta_2}\}$,

$P_2=\{\ketbra{\theta_1}{\theta_1},\ketbra{\theta_2}{\theta_2},\ketbra{\theta_0}{\theta_0}\}$,

$P_3=\{\ketbra{\theta_2}{\theta_2},\ketbra{\theta_0}{\theta_0},\ketbra{\theta_1}{\theta_1}\}$.

According to \cite{virmani2003construction}, the POVM $P_i$ are all extremal POVMs. Note that, when the basis of $\theta$ is equal to the computational basis, it corresponds to the particular case, $\text{POVM}_{\{E_{B^{m}}^{l}\}}=\frac{1+2\varepsilon_{m}}{3}P_1+\frac{1-\varepsilon_{m}}{3}P_2+\frac{1-\varepsilon_{m}}{3}P_3$.  
\par

Although we are uncertain if there exist any other forms of extreme value decompositions for the POVM, based on numerical comparisons in device-independent scenarios introduced in the subsequent section, we can infer that this form may provide the maximum guessing probability, or, to be more precise, yield an approximate strategy that generates a guessing probability close to the maximum. 
\revised{
Since it is challenging for existing methods to explore all extremal decomposition forms of POVMs under high-dimensional systems in the device-dependent scheme, we choose a special extremal decomposition form of POVMs (equation (\ref{extremal_de})) to quantify the quantum intrinsic randomness under the CGLMP inequality. This decomposition form may not yield the maximum guess probability in the device-dependent scheme. More precisely, our POVM decomposition form is only an approximation strategy for the maximum guess probability in this scheme. In figure \ref{fig4} and \ref{fig5}, we compare the randomness generated by these two schemes. The randomness between them is relatively close. Even if there are other POVM decomposition strategies, the maximum guess probability obtained by them will be around equation (\ref{extremal_de}), but it will not exceed the maximum guess probability in the device-independent scheme.}
% Even if there are other strategies for POVM decomposition, the guessing probability, in that case, would still be close to the one with equation (\ref{extremal_de}) and not exceed the one in device independence (see the randomness comparison in figure \ref{fig4} and figure \ref{fig5} in the next subsection). 
In the device-independent scenario, we assume both the state and measurement are untrusted, and employ a numerical tool to explore all possible decomposition forms, thereby providing an upper bound of the guessing probability.

\begin{figure}[H]
\centerline{\includegraphics[width=0.7\linewidth]{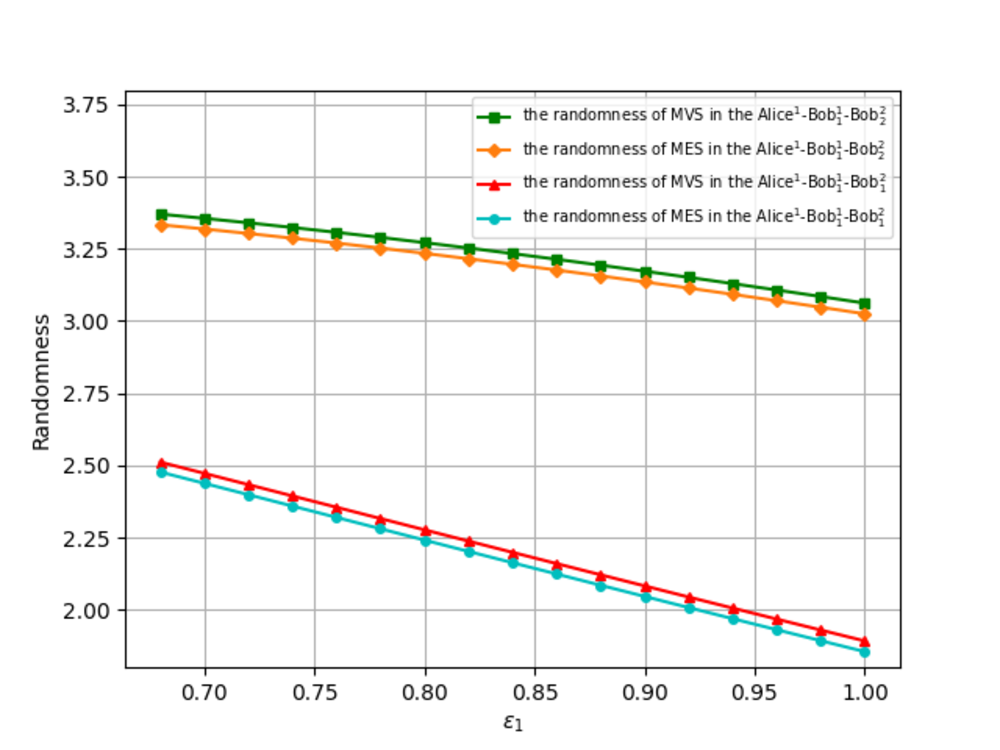}}
\caption{The global randomness in sequential CGLMP scenario with MVS and MES. The horizontal axis represents the range in which double CGLMP inequality violation occurs. Global randomness refers to the logical expression of the negative guessing probability as described in equation (\ref{global guess probability}).}
\label{fig3}
\end{figure}

We begin by considering the maximum entanglement state in the trusted quantum randomness system scenario, we can get the quantum randomness for a given measurement setting.
 Except for the local randomness given in the next section in figure \ref{fig4}, local randomness refers to the randomness observed in the joint outcomes of variables $b_1$ and $b_2$. 
We give the Alice Bob$^{1}$ and Bob$^{2}$ global randomness expression. Global randomness here refers to the randomness observed in the joint outcomes of variables $a$, $b_1$, and $b_2$. Additionally, we conducted a randomness analysis specifically for the maximum violation state.
Subsequently, we present a comparison of the global intrinsic randomness between the maximum violation state and the maximum entangled state, as illustrated in figure \ref{fig3}. This comparison is based on the measurement decomposition outlined in the equation
(\ref{extremal_de}). The results indicate that in the sequential CGLMP measurement scenario, the maximum violation state exhibits a higher degree of randomness compared to the maximum entangled state. Furthermore, the randomness in the Alice$^1$-Bob$_1^1$ -Bob$_2^2$ measurement is higher than that in the Alice$^1$-Bob$_1^1$ -Bob$_1^2$ measurement, where the superscript denotes the ordinal number of the participant's measurement. It should be noted that only the randomness in Alice$^1$-Bob$_1^1$ -Bob$_2^2$ and Alice$^1$-Bob$_1^1$ -Bob$^2_1$ measurements are showcased here, but due to symmetry, the same holds for the randomness in other measurement outcomes as well.
% \begin{figure}[h]
% \centerline{\includegraphics[width=0.8\linewidth]{mvs_mes_white_randomness.png}}
% \caption{MVS and MES Comparison of randomness in white noisy.}
% \label{fig}
% \end{figure}

\subsubsection{quantum randomness in device independent scenario} 
\revised{We have analyzed the randomness within the CGLMP framework under the condition of a trusted quantum system, where the form of each Bob's POVM is known but lacks decomposition information. Subsequently, we will investigate the randomness in the context of untrusted devices, demonstrating how trustworthiness can impact quantifiable randomness. By comparing the randomness in trustworthy devices without decomposition information to that in untrusted device scenarios (device-independent) with NPA quantification, we aim to provide a comprehensive understanding of randomness across different levels of trustworthiness, thereby enriching our research.}

In device-independent quantum random generators, the quantum systems are regarded as black boxes, and no assumptions are made about the physical dimension of the underlying state. Due to the one-to-many mapping between classical statistics and quantum systems, there can be multiple realizations of the state and measurements that yield the same statistics. Among these realizations, some may provide Eve with an advantage in obtaining information about the observed outcomes. Therefore, to quantify the randomness generated, it is necessary to consider all possible realizations and determine the lower bound on the amount of randomness that can be certified from an observed probability distribution in the device-independent setting.

However, considering all possible realizations of the statistics poses significant computational challenges, especially when the dimension of the underlying Hilbert space is unknown.  To resolve this technical difficulty,  Navascu\'es, Pironio, and Ac\'in \cite{navascues2007bounding} transformed the problem of verifying quantum correlations into a semidefinite feasibility problem by introducing linear constraints and semidefinite constraints. This technique, commonly referred to as the NPA method, allows for effective resolution using semidefinite programming algorithms.  Building upon the original NPA hierarchy, Bowles et al. \cite{bowles2020bounding} further enhanced the method by incorporating additional sequential constraints, giving rise to the construction of the sequential NPA hierarchy. In this section, we leverage the sequential NPA hierarchy to establish bounds on the amount of randomness present within observed probability distributions,  utilizing full probability as a constraint. Rather than depending on violation inequalities, the use of full probabilities as constraints is preferred due to its demonstrated optimality in quantifying randomness \cite{bancal2014more,nieto2014using}.\par
\revised{The randomness defined is not suitable for untrusted devices anymore, since the state and measurements form are unknown now. We begin by introducing Eve's guessing probability as the measure of randomness and then using NPA methods to provide an upper bound about the maximum guessing probability.} 
The guessing probability for Alice's input $ x = x^{\ast}$ and Bob’s input $ y = y^{\ast}$, denoted as $ G(x^{\ast},y^{\ast})$, refers to the maximum probability that Eve can correctly guess both Alice and Bob's outputs while reproducing the observed probability distribution $p_{AB}(a,b_{1},b_{2}|x,y_{1},y_{2})$ 
 when marginalizing her output. Mathematically, it is defined as follows:
%The  guessing probability for Alice's input $ x = x^{\ast}$ and Bob’s input $ y = y^{\ast}$ given an observed probability distribution
%$P_{\text{obs}}(a, b|x, y)$ is the best probability that Eve can guess Alice and Bob's output, while also reproducing $P_{\text{obs}}$ when marginalizing over her output. That is,
\begin{equation}
    G(x^{\ast},y^{\ast}) = \max_{ p_{ABE}\in Q}\sum_{e=a,b_1,b_2}p(a,b_1,b_2,e|x^{\ast},y_{1}^{\ast},y_{2}^{\ast}). \label{global_guess}
\end{equation}
Here, $p_{ABE}$ represents the joint probability distribution of Alice, Bob, and Eve's outputs, constrained within the quantum set 
$Q$, which is the set that contains all the quantum realizations. The sum is taken over the possible outcomes 
$e$ for Eve, which includes the alphabets of Alice and Bob's outputs, namely 
$a, b_1$, and $b_2$. Here $e$ is used to represent Eve's guessing outcome of the alphabet of Alice and Bob’s outputs and its size is determined by the product of the sizes of $a, b_1$, and $b_2$. It should be noted that we assume Eve has no input, as she can use a single measurement that has the size of $|a|\cdot|b_1|\cdot|b_2|$ outputs to guess the outcomes. The observed probability distribution $p_{AB}(a,b_{1},b_{2}|x,y_{1},y_{2})$
 is related to the joint distribution $p_{ABE}(a,b_{1},b_{2},e|x,y_{1},y_{2})$
through the following formula:
\begin{equation}
\begin{split}
    p_{AB}(a,b_{1},b_{2}|x,y_{1},y_{2}) &= \sum_{e}p_{ABE}(a,b_{1},b_{2},e|x,y_{1},y_{2})\\
    &= p_{\text{obs}}(a,b_{1},b_{2}|x,y_{1},y_{2}).
\end{split}
\end{equation}
\revised{To obtain an observable probability distribution $p_{\text{obs}}$, we also consider the above scenario of one Alice and a sequence of two Bobs, initially sharing the maximum entangled state and the maximum violation state.} 
In the device-independent scenario, the underlying POVM taken by all participants as well as the initial shared state between Alice and Bob$^1$ are untrustworthy. Therefore, the observed probabilistic statistics can originate from any quantum subsystem in any dimension. Since mixed states and POVM can be extended to pure states and projection systems in higher dimensions, we only focus on the implementation of pure state and projection measurements. Therefore, the observed probability has the following quantum realization,
\begin{equation}
\begin{split}
    p_{AB}(a,b_{1},b_{2}|x,y_{1},y_{2}) &= \la\psi|\Pi_x^a\otimes\Pi_{y_1}^{b_1}\Pi_{y_2}^{b_2}|\psi\ra.
\end{split}
\end{equation}
As the Naimark extension of the POVM introduced in figure \ref{figsequentialM}, $\Pi_{y_1}^{b_1}\Pi_{y_2}^{b_2}$ may have the form,
$\Pi_{y_1}^{b_1}\Pi_{y_2}^{b_2} = U_{y_1}^\dagger U_{y_2}^\dagger (I\otimes \Pi_{b_1}\otimes \Pi_{b_2})
U_{y_1}U_{y_2}$. The sequential NPA hierarchy introduced by Bowles et al is used to run over all the possible pure state $|\psi\rangle$ and projective measurement $\Pi_{y_1}^{b_1}\Pi_{y_2}^{b_2}$ to upper bound the guessing probability of equation (\ref{global_guess}).

Before that, it is necessary to provide the calculation of the observed statistics in the specified experiment. The states under consideration are both maximum entanglement states and maximum violation states.  The POVM  in Bob$^1$ is defined as given in equation (\ref{white-noisy measurement}), and we assume it can be decomposed into,
\begin{equation} 
\begin{split}
 \revised{E_{B^{m}_y}^{l}} &= \frac{1+2\varepsilon_{1}}{3}|l\rangle_{B}\langle l|_{B}+\frac{1-\varepsilon_{1}}{3}|l+1 \bmod 3\rangle_{B}\langle l+1 \bmod 3|_{B}\\
 &+ \frac{1-\varepsilon_{1}}{3}|l+2 \bmod 3\rangle_{B}\langle l+2 \bmod 3|_{B}.
\end{split}\label{NPA_decom}
\end{equation}
\revised{
In equation (\ref{white-noisy measurement}), $I_{3}=|l\rangle _{B}\langle l|_{B}+|l+1 \mod 3 \rangle _{B}\langle l+1 \mod 3|_{B}+|l+2 \mod 3 \rangle _{B}\langle l+2 \mod 3|_{B}$, $|l\rangle_{B}$ is expressed by equation (\ref{Cglmp_b}).
When |$\theta_0$⟩, |$\theta_1$⟩, |$\theta_2$⟩ in equation (\ref{extremal_de}) are computational basis,
equation (\ref{NPA_decom}) is a special form of equation (\ref{extremal_de}). 
We denote the coefficient and the operators as $\varepsilon_1$ and $\{E_{B^{m}_y}^{l}\}_{\varepsilon_1}$. The range of values for the coefficient $\epsilon_1$ is the range in which the above sequential CGLMP scenario can achieve double violations, then the observed statistics can be calculated as follows:
\begin{equation}  G(a,\vec{b}|x,\vec{y})= \sum_{l}\tr(\rho_{AB}\{E_{B^{m}_y}^{l}\}_{\varepsilon_1}) \tr(\{E_{B^{m}_y}^{l}\}_{\varepsilon_1}\rho_{AB}\{E_{B^{m}_y}^{l}\}_{\varepsilon_1}^\dagger M_{x}^{a}\otimes M_{y_2}^{b_2})).  
\end{equation}
}The quantum randomness for the given measurement setting $(x,\vec{y})$then is quantified by
\begin{equation}
   H_{\infty}(a\vec{b}|x\vec{y}) = -\log_2 G(a,\vec{b}|x,\vec{y}).
\end{equation}
We utilize the 1+AB level of the sequential NPA algorithm to quantify the randomness in the sequence CGLMP  setup. Given that both mixed states and POVM can be extended to pure states and projection systems in higher dimensions, we focus solely on the implementation of pure state and projection measurements. Consequently, it is necessary to introduce 6 operators each to characterize the measurements of Alice, Bob$^1$, and Bob$^2$. The $AB$ moment entails a total of 98 operators, whereas there exist 36 operators representing the product of operators between Bob$^1$ and Bob$^2$.  Hence, the resulting matrix dimension is $108\times108$. Additionally, we need to consider introducing operators for Eve to characterize Eve's guessing outcomes. Given the practical scale of the NPA hierarchy for sequences, we limit our examination to local randomness. This approach eliminates the need to describe Eve using 27 projection operators to guess global outcomes within the algorithm, opting instead for just 9 operators to guess two Bobs' outcomes, thereby improving computational efficiency. Although the sequential NPA allows for the theoretical characterization of global randomness, the algorithm significantly expands in size, leading to slower computational speed.
\begin{figure}[H]
\centerline{\includegraphics[width=0.7\linewidth]{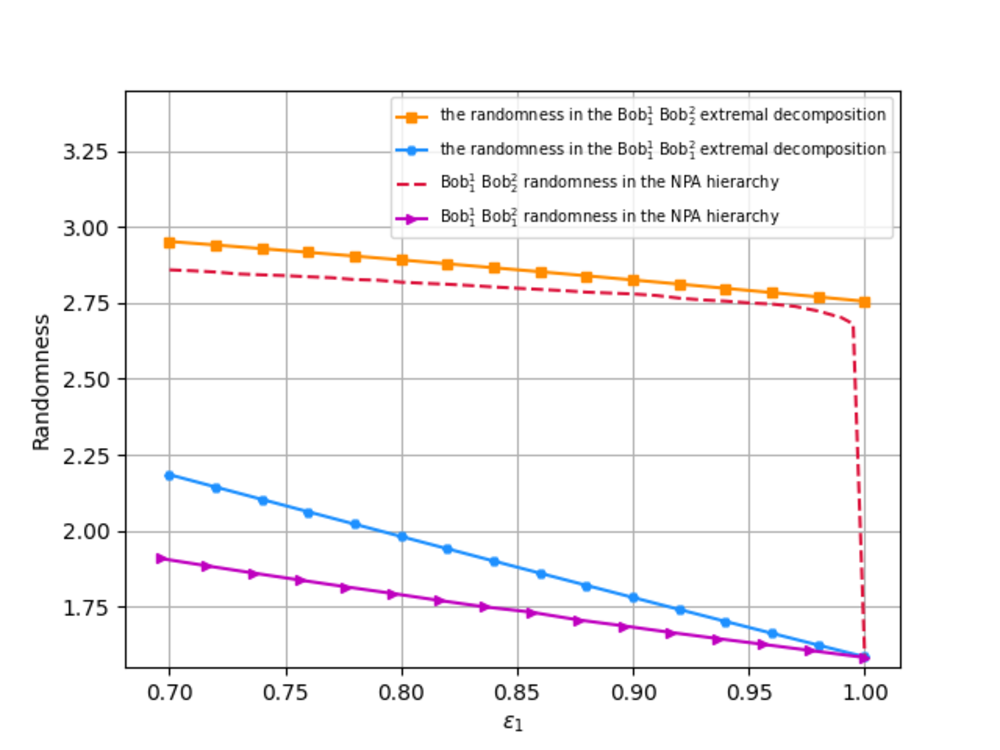}}
\caption{ The randomness in the MES,  with the given extreme value decomposition of  POVM in equation (\ref{extremal_de}) and the sequential NPA hierarchy approach.}
\label{fig4}
\end{figure}

\begin{figure}[H]
\centerline{\includegraphics[width=0.7\linewidth]{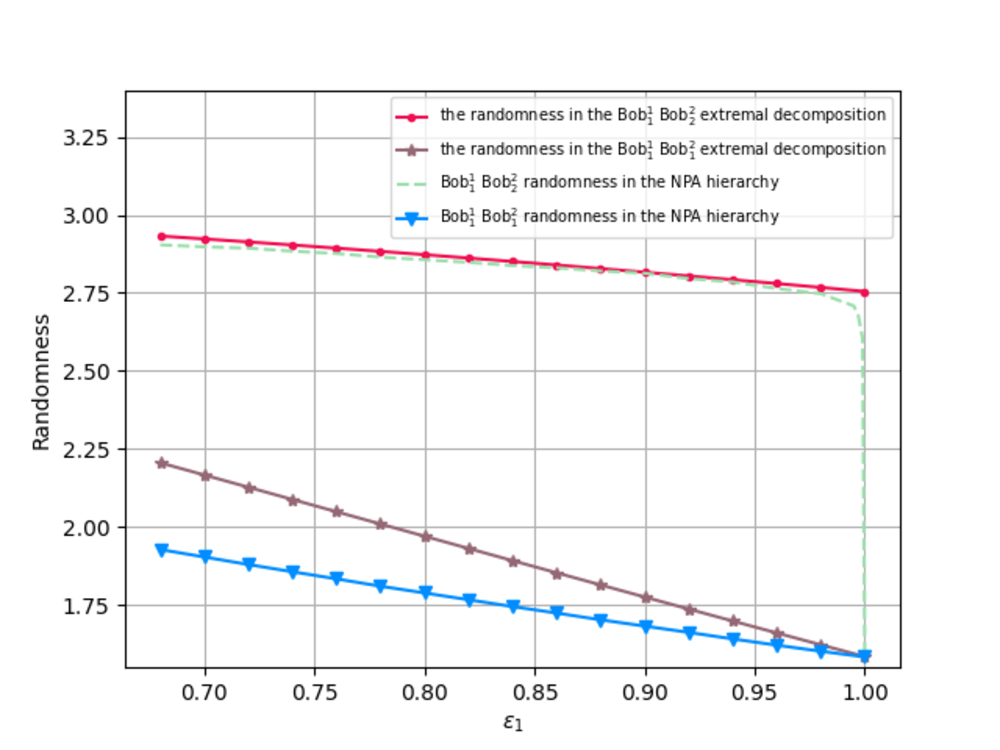}}
\caption{The randomness in the MVS, with the given extreme value decomposition of POVM in equation (\ref{extremal_de}) and the sequential NPA hierarchy approach.}
\label{fig5}
\end{figure}

In figure \ref{fig4} and figure \ref{fig5},  we present lower bounds to the logarithm of the negative about $G(y^{*} = (0, 1))$ and $G(y^{*} = (0, 0))$, namely randomness,  calculated using level 1 + AB of the hierarchy, taking into account the effect of noise. In figure \ref{fig4} the observed statistics are based on the maximum entanglement state, while figure \ref{fig5} is based on the maximum violation state.
We compared the randomness achieved by different measurement bases in figures \ref{fig4} and \ref{fig5}. The results show that the randomness corresponding to different measurements chosen by Bob$^1$ and Bob$^2$ is different. In particular, the randomness of Bob$_{1}^{1}$ and Bob$_{2}^{2}$ consistently exceeds that of Bob$_{1}^{1}$ and Bob$_{1}^{2}$, regardless of whether it is the maximum entangled state or the maximum violation state.

In figures \ref{fig4} and \ref{fig5}, within the device-trusted scheme, the extreme decomposition form of equation (\ref{extremal_de}) is used to derive the randomness associated with the maximum entangled state and the maximum violation state. In addition, in the device-independent scenario, the sequential NPA hierarchy is used to quantify randomness. For the setting of Bob$_{1}^{1}$ and Bob$_{2}^{2}$, it shows that the randomness coming from these two methods is quite close enough, suggesting the randomness for trusted scenario is quite close to the one from our given decomposition (\ref{extremal_de}). For the setting of Bob$_{1}^{1}$ and Bob$_{1}^{2}$, the randomness coming from those two methods is relatively larger,  thus we infer there may be more optimal decomposition for the setting of Bob$_{1}^{1}$ and Bob$_{1}^{2}$ or the NPA hierarchy for the setting of Bob$_{1}^{1}$ and Bob$_{1}^{2}$ is not tight enough.

\section{Conclusion}
In this paper, we present a definition of quantum intrinsic randomness under sequential measurements. To achieve this goal, 
Eve's classical maximum guess probability and the quantum maximum guess probability are provided based on the varying degrees of association between the eavesdropper Eve and the quantum system. 
Furthermore, we have quantified the quantum intrinsic randomness under the CGLMP sequential measurements scenario at various confidence levels in the device. In the trusted quantum system scenario, the quantum state is pure, maximum entanglement state or maximum violation state, and only the measurement is noisy, in which case Eve has the same maximum quantum guess probability and maximum classical guess probability, so only POVM extremal decomposition should be considered to quantify the classical guess probability. For POVM in the CGLMP scenario, we provide a special kind of decomposition and calculate the randomness in the result of this decomposition. This special kind of decomposition provides an upper-bound estimate of randomness. In principle, we should run over all the extremal decomposition of POVM to quantify randomness. However, providing all decomposition forms is challenging, especially for high-dimensional POVM. In future studies, we will further explore high-dimensional POVM extremum decomposition methods to obtain more accurate results. In the scenario of untrusted quantum systems, both quantum states and measurements may contain noise, we use the sequential NPA hierarchy method to lower the bound of the verifiable quantifiable randomness. In any case,
our work imposes stricter constraints on the definition of intrinsic randomness in sequential measurements. It has the potential to help improve random number generation rates in practice.

Our research contributes to understanding the limitations and possibilities of generating random numbers in sequential measurements in quantum systems. The defined quantum intrinsic randomness provides a basis for further exploration and applications in quantum information processing and cryptography. In the future, it is possible to extend our results to more non-locality scenarios, including the two-sided sequential measurement scenario, semi-device steering scenario \cite{yukun2023}, and probability statistic criteria as non-locality scenarios \cite{yukun2016}. 
It remains to be investigated whether the relationship between sequential intrinsic randomness and sequential nonlocality and entanglement is consistent with the standard Bell scenario.

\section*{Acknowledgement}

This research was supported by the National Nature Science Foundation of China (Grants No. 62101600, No. 62201252), the Science Foundation of China University of Petroleum, Beijing (Grant No. 2462021YJRC008), State Key Laboratory of Cryptology (Grant No. MMKFKT202109), and Natural Science Foundation of Jiangsu Province, China (Grant No. BK20190407).\\

\section*{Appendix}
\appendix
We give the proofs of two theorems in this section.
\textbf{The proof of Theorem 1}
\begin{proof}

Let $\sum_{\lambda}p_{\lambda}| \phi_{\lambda}\ra |e_{\lambda}\ra$ be an purification of $\rho_S$.  We could define the measurement \revised{$\{\Pi_E^{\vec{b}}\}_{\vec{b}}$} on Eve’s
subsystem. \revised{$\{\Pi_E^{\vec{b}}\}_{\vec{b}}$}  will project onto the subspace spanned by the states $\lambda$ such that $\vec{b}$ maximizes the Born rule for
the $i$-th state of $\rho_S$. That is,
\revised{
$$\Pi_E^{\vec{b}}=\sum_{\lambda \in A_{\vec{b}}}|e_\lambda \ra \la e_\lambda| $$ }
with
\revised{
 $A_{\vec{b}}= \{\lambda|\vec{b}= 
 \min \{ \vec{x}| \la \phi_{\lambda}| \Pi_{\vec{y}}^{\vec{x}}(\Pi_{\vec{y}}^{\vec{x}})^\dagger|\phi_{\lambda}\ra =\max_{\vec{z}} \la \phi_{\lambda}| \Pi_{\vec{y}}^{\vec{z}}(\Pi_{\vec{y}}^{\vec{z}})^\dagger\phi_{\lambda}\ra\} \}$, where we have denote $\Pi_{\vec{y}}^{\vec{x}} := \Pi_{y_1}^{x_1}...\Pi_{y_n}^{x_n}$.}
 
Suppose \revised{$\{\Pi_E^{\vec{b}},|\phi_{SE}\}$} is a solution to equation (\ref{Q_guess_noisestate}), and for every solution we have,
\revised{
\begin{equation}
\begin{split}
   \sum_{\vec{b}}&\sum_{\lambda \in A_{\vec{b}}} \la \psi_{SE}| \Pi_{y_1}^{b_1}...\Pi_{y_n}^{b_n}|e_\lambda \ra \la e_\lambda| \Pi_{y_n}^{b_n}...\Pi_{y_1}^{b_1}|\psi_{SE}\ra\\
    &=\sum_{\lambda }\max_{\vec{b}} \la \phi_{\lambda}| \Pi_{y_1}^{b_1}...\Pi_{y_n}^{b_n}\Pi_{y_{n-1}}^{b_{n-1}}...\Pi_{y_1}^{b_1}|\phi_{\lambda}\ra\\
    &=p^C_{\text{guess}}(\vec{b}|\vec{y},\rho_s,\{\Pi_{y_i}^{b_i}\}_i,E)
\end{split}
\end{equation}}
 Therefore,  \revised{$$p^Q_{\text{guess}}(\vec{b}|\vec{y},\rho_s,\{\Pi_{y_i}^{b_i}\}_i,E) = p^C_{\text{guess}}(\vec{b}|\vec{y},\rho_s,\{\Pi_{y_i}^{b_i}\}_i,E)$$}
\end{proof}
\textbf{The proof of Theorem 2}.
\begin{proof}
\textbf{Firstly, there is a fact that $$p^Q_{\text{guess}}(\vec{b}|\vec{y},|\psi_S\ra,\{M_{S}^{b_i}\}_i,E)\geq p^C_{\text{guess}}(\vec{b}|\vec{y},|\psi_S\ra,\{M_{S}^{b_i}\}_i,E)$$}

Let $${(p(j), \{M_S
^{b_i,j}\}_{b_i}, \ket{\psi_S}}$$ be an optimal solution to $p^C_{\text{guess}}(\vec{b}|\vec{y},|\psi_S\ra,\{M_{S}^{b_i}\}_i,E)$.     Consider a bipartite ancillary
system $A_i = A'_iA''_i$ initially in the state
$$\sigma_{A_i} = \ketbra{0}{0}_{ A'_i}\otimes \sum_{j}p(j)\ketbra{j}{j}_{A''_i}$$
with $\text{dim}(H_{A'_i}
) = |d_i|$. Let Eve hold the purification
\begin{align}
  |\phi_{A_1,...,A_nE}\ra =\sum_{\textbf{j}}\sqrt{p(\textbf{j})}
\ket{\textbf{0},\textbf{j}}_{A'A''}
\ket{\textbf{j}}_E  
\end{align}\label{phi_AE}
where $\textbf{0}$ represents $n$ initial state $0$ in each $A_i'$ side and $\textbf{j}=j_1,...,j_n$, each $j_i$ is in $A_i''$ side. 
Notice that $ \tr_{A_{j\neq i}E}[\ketbra{\phi}{\phi}_{\textbf{A}E}] = \sigma_{A_1}$, where we denote $A_1,...,A_n$ as $\textbf{A}$. Define an operator $U_{y_i}$ via its action on the state $\ket{\psi_S}\ket{\textbf{0},\textbf{j}}_{A'_iA''_i} $ as

$$U_{y_i}\ket{\psi_S}\ket{\textbf{0},\textbf{j}}_{A'_iA''_i}
=\sum_{b_i}\sqrt{M_{y_i}^{b_i,j_i}}\ket{\psi,b_i,j_i}_{SA'_iA''_i}\ket{\textbf{j}/i}
$$
 It follows that $U_{y_i}$
can be extended to
a unitary operator acting on $\ket{\psi_S}\ket{0,j_i}_{A'_iA''_i} $.

With the first $i$ unitary $U_{y_i}$ we have
\begin{align}
&U_{y_i},...,U_{y_1}\ket{\psi_S}\ket{\textbf{0},\textbf{j}}_{\textbf{A}}\\\nn
&=\sum_{b_1,...,b_i}\sqrt{M_{y_i}^{b_i,j_i},...,M_{y_1}^{b_1,j_1}}\ket{\psi,b_1,...,b_i,j_1,...,j_i}_{S\textbf{A}}\ket{j_1,...,j_i}_E\\ \nn
& \ket{0,\textbf{j}/\{1,...,i\}}_{\textbf{A}/\{1,...,i\}} \ket{\textbf{j}/\{1,...,i\}}_E 
\end{align}

Measure the $A'_i$
space in the $\ket{b_i}$ basis, obtaining outcome $b_i$.
Conditioning on outcome $b_i$ and the first $i-1$ outcomes $b_{i-1},...,b_1$, then tracing out  $\textbf{A}$ spaces, one finds 
\begin{align}
&\tr_{\textbf{A}}({\sum_{\textbf{j}}\sqrt{p(\textbf{j})}U_{y_i},...,U_{y_1}\ket{\psi_S}\ket{\textbf{0},\textbf{j}}_{\textbf{A}}}{\sum_{\textbf{j}}\sqrt{p(\textbf{j})}{U_{y_1},...,U_{y_i}}\bra{\psi_S}\bra{\textbf{0},\textbf{j}}_{\textbf{A}}}) \nn \\ 
&=\sum_{j_1,...,j_i}{p(j_1,...,j_i)}\sqrt{M_{y_i}^{b_i,j_i},...,M_{y_1}^{b_1,j_1}}\ketbra{\psi}{\psi}\sqrt{M_{y_i}^{b_1,j_1},...,M_{y_i}^{b_i,j_i}} \nn \\
&:=\rho_{\text{post}}(b_1,...,b_i|y_1,...,y_i)   
\end{align}

 It's corresponding
post-measurement state described by the first $i$ POVM, with a decomposition of a set of Kraus operators $\{M_{y_i}^{b_i,j_i}\}_{j_i}$.
We have thus
reproduced the $i$-th measurement with high-dimension projectors.

$$\Pi_{y_i}^{b_i} = U_{y_i}^\dagger (I\otimes \Pi_{b_i})
U_{y_i}$$

Notice that, 
$$\tr_{A_i}[\Pi_{y_i}^{b_i} I_S \otimes \sigma_{A'_iA''_i}] =  M_{y_i}^{b_i}$$
therefore, $\{\Pi_{y_i}^{b_i}, \sigma_{A'_iA''_i}\} $gives a projective extension of $ \{ M_{y_i}^{b_i}\}$.
 Repeating this for the  measurement in the sequence, we have  a sequence of arbitrary length,%we find
%$$\Pi_{y_1}^{b_1}\Pi_{y_2}^{b_2} = U_{y_1}^\dagger U_{y_2}^\dagger (I\otimes \Pi_{b_1}\otimes \Pi_{b_2})
%U_{y_1}U_{y_2}$$
%This processing can be continued, then 
 $$\Pi_{y_1}^{b_1}...\Pi_{y_n}^{b_n} = U_{y_1}^\dagger ...U_{y_n}^\dagger (I\otimes \Pi_{b_1}\otimes... \otimes \Pi_{b_n})
U_{y_1}...U_{y_2}$$
 be projective without loss of generality.
 
 We could define the measurement $\{M_E^{\vec{e}}\}_{\vec{e}}:=M_{E_1}^{e_1}\otimes...\otimes M_{E_n}^{e_n} $ on Eve’s
subsystem. Notice that, each Eve can cooperate, the measurements of $i$-th Eve depend on the strategies of all the previous $i-1$ Eve. Thus, instead of considering local Eve's measurement, we introduce a joint measurement (as figure \ref{figsequentialM} shown, correlations may exist between different Eves.)
$\{M_{\vec{E}}^{{\vec{b}}}\}_{{\vec{b}}}$ in  Eve, which will project onto the space spanned by the states $\ket{j_1,..,j_n}$ such that ${b_1,..,b_n}$ maximizes the Born rule for
the $j_i$-th POVM decomposition of $M_{y_i}^{b_i}$ acting on the post-measurement state obtained by first $i-1$ measurements. That is,
$$M_{\vec{E}}^{{\vec{b}}}=\sum_{j \in A_{{\vec{b}}}}|\textbf{j}  \ra \la \textbf{j} | $$ 
with
 $A_{{\vec{b}}}= \{\textbf{j}|{\vec{b}}= 
 \min \{ \vec{x}| \la \psi_{S}| (M_{{y_1}...{y_{i-1}}}^{{x_1}...{x_{i-1}}} )^\dagger M_{{y_i}}^{{x_i,j_i}} (M_{{y_1}...{y_{i-1}}}^{{x_1}...{x_{i-1}}} )|\psi_{S}\ra =\max_{\vec{z}} \la \psi_{S}| (M_{{y_1}...{y_{i-1}}}^{{z_1}...{z_{i-1}}} )^\dagger M_{{y_i}}^{{z_i,j_i}} M_{{y_1}...{y_{i-1}}}^{{z_1}...{z_{i-1}}} |\psi_{S}\ra \}$, where we have denote $M_{{y_1}...{y_{i-1}}}^{{b_1}...{b_{i-1}}} :=\sqrt{M_{y_1}^{b_1,j_1}}...\sqrt{M_{y_{i-1}}^{b_{i-1},j_{i-1}}}$, $\vec{x}=x_1,...,x_n$ and $\vec{z}=z_1,...,z_n$ . 
 
Suppose  $\{M_{\vec{E}}^{\vec{b}},| \phi_{A_1E_1}\ra... | \phi_{A_1E_1}\ra \}$ is a solution to equation (\ref{Q_guess_noiseM}), and for every solution we have,
 \begin{equation}
\begin{split}
   \sum_{\vec{b}} &\la \psi_S| \la \phi_{A_1E_1}|... \la \phi_{A_nE_n}|{\Pi_{y_1}^{b_1}}...\Pi_{y_n}^{b_n}\\
&\otimes M_{\vec{E}}^{\vec{b}}\otimes{\Pi_{y_n}^{b_n}}...\Pi_{y_1}^{b_1}
 | \phi_{A_1E_1}\ra... | \phi_{A_1E_1}\ra |\psi_S\ra \\
 &=\sum_{\vec{b}} \la \psi_S| \la \phi_{A_1E_1}|... \la \phi_{A_nE_n}|{\Pi_{y_1}^{b_1}}...\Pi_{y_n}^{b_n}\\
&\otimes \sum_{j_1,...,j_n \in A_{{\vec{b}}}}|j_1,...,j_n  \ra \la j_1,...,j_n |...\otimes 
 | \phi_{A_1E_1}\ra... | \phi_{A_1E_1}\ra |\psi_S\ra \\
    &=\sum_{j_1,...,j_n}p_{j_1}\cdot...\cdot p_{j_n}\max_{b_1,...,b_n} \la \psi_{S}| \sqrt{M_{y_1}^{b_1,j_1}}^\dagger...\sqrt{M_{y_{n-1}}^{b_{n-1},j_{n-1}}}^\dagger \\
    &\cdot M_{y_{n}}^{b_{n},j_n} \sqrt{M_{y_{n-1}}^{b_{n-1},j_{n-1}}}...\sqrt{M_{y_1}^{b_1,j_1}} |\psi_{S}\ra\\
    &=p^C_{\text{guess}}(\vec{b}|\vec{y},\rho_s,\{M_{y_i}^{b_i}\}_i,E)
\end{split}
\end{equation}
 Therefore, $$p^Q_{\text{guess}}(\vec{b}|\vec{y},|\psi_S\ra,\{M_{S}^{b_i}\}_i,E)\geq p^C_{\text{guess}}(\vec{b}|\vec{y},|\psi_S\ra,\{M_{S}^{b_i}\}_i,E)$$

\textbf{Secondly, for pure initial state  case, there is 
{$$p^Q_{\text{guess}}(\vec{b}|\vec{y},|\psi_S\ra,\{M_{S}^{b_i}\}_i,E)\leq p^C_{\text{guess}}(\vec{b}|\vec{y},|\psi_S\ra,\{M_{S}^{b_i}\}_i,E)$$}}
 This comes from the fact that for every quantum solution,denoted as $ \{\Pi_{y_i}^{b_i},M_{\vec{E}}^{\vec{x}},\ket{\psi_S}| \phi_{A_1E_1}\ra... \ket{\psi_S}| \phi_{A_nE_n}\ra \}$,  to $p^Q_{\text{guess}}(\vec{b}|\vec{y},|\psi_S\ra,\{M_{S}^{b_i}\}_i,E)$, there corresponding to a decomposition of $M_{y_i}^{b_i}$ as $M_{y_i}^{b_i,(x_i,j_i)}=\tr_{A_i}[\Pi_{y_i}^{b_i}(I_S\otimes \ketbra{\psi_{j_i}^{x_i}}{\psi_{j_i}^{x_i}}]$ which gives a solution to $p^C_{\text{guess}}(\vec{b}|\vec{y},|\psi_S\ra,\{M_{S}^{b_i}\}_i,E)$.

 Concretely, the post-measurement states on $H_{SA_iE_i}$
are all separable between the system S and the system A since the initial state $\ket{\psi_S}$ is a pure state. We denote the post-measurement states on $H_{SA_iE_i}$ as 
$$\tau^{x_1,...,x_n}_
{SA_i} = \frac{
\tr_{E_i}[(I_S \otimes I_{A_i} \otimes M^{x_1,...,x_n}_{\vec{E}}\ketbra{\psi}{\psi}_{SAE}]}{p(x_1,...,x_n)}
$$

with
$p(x_1,...,x_n) = \tr[(I_S\otimes I_{\textbf{A}} \otimes M^{x_1,...,x_n}_{\vec{E}}\ketbra{\psi}{\psi}_{SAE}]$, where
${\psi}_{SAE}=\ket{\psi_S}\ket{\phi_{AE}}$. The $\ket{\phi_{AE}}$ is as given in \ref{phi_AE}.
We write its separable form as, 
$\tau^{x_1,...,x_n}_
{SA_i} = \ketbra{\psi}{\psi}_S\otimes\ketbra{\psi^{\vec{x}}}{\psi^{\vec{x}}}_{\textbf{A}}
$, by denoting $\ket{\textbf{0},x_1,...,x_n}$ as $\ket{\psi^{\vec{x}}}$.

And define $M_{y_i}^{b_i,\vec{x}}=\tr_{A_i}[\Pi_{y_i}^{b_i}(I_S\otimes \ketbra{\psi^{\vec{x}}}{\psi^{\vec{x}}}_{A}]$.

Notice that, 
 \begin{equation}
\begin{split}\sum_{\vec{x}}&p(\vec{x},j)M_{y_i}^{b_i,\vec{x}}=\sum_{\vec{x}}p(\vec{x})\tr_{A_i}[\Pi_{y_i}^{b_i}(I_S\otimes \ketbra{\psi^{\vec{x}}}{\psi^{\vec{x}}}_{A_i}]\\
&=\tr_{A_i}[\Pi_{y_i}^{b_i}(I_S\otimes \sum_{\vec{x}}p(\vec{x})\ketbra{\psi^{\vec{x}}}{\psi^{\vec{x}}}_{A_i}]\\
&=\tr_{A_i}[\Pi_{y_i}^{b_i}(I_S\otimes \sigma_{A_i}]=M_{y_i}^{b_i}
\end{split}
 \end{equation}
and that $M_{y_i}^{b_i,\vec{x}}\geq0$,  satisfying$\sum_{b_i}M_{y_i}^{b_i,\vec{x}}=I_s$
Therefore, $\{(p(\vec{x}), \ket{\psi_S}, M_{y_i}^{b_i,\vec{x}}\}$ is a solution to equation (\ref{pguess_C_noiseM}) with value
\begin{equation}
\begin{split}
\sum_{\vec{x}}&\max_{b_1,...,b_n}\la \psi_S|\cdot\sqrt{M_{y_1}^{b_1,x_1}}^\dagger...M_{y_n}^{b_n,x_n}...\sqrt{M_{y_1}^{b_1,x_1}}|\psi_S\ra\\
&=\sum_{\vec{x}}p(\vec{x})\max_{b_1,...,b_n} \la \psi_S| \la \phi_{A_1E_1}|... \la \phi_{A_nE_n}|\\
&\cdot (\Pi_{\vec{y}}^{\vec{b}})^\dagger\otimes M_{{E}}^{{x_1,...,x_n}}\otimes \Pi_{\vec{y}}^{\vec{b}} 
 | \phi_{A_1E_1}\ra... | \phi_{A_1E_1}\ra |\psi_S\ra \\
&\geq \sum_{\vec{x}}p(\vec{x})\la \psi_S| \la \phi_{A_1E_1}|... \la \phi_{A_nE_n}|\\
&\cdot (\Pi_{\vec{y}}^{\vec{x}})^\dagger\otimes M_{E}^{{x_1,...,x_n}}\otimes\Pi_{\vec{y}}^{\vec{x}} 
 | \phi_{A_1E_1}\ra... | \phi_{A_1E_1}\ra |\psi_S\ra \\
&=p^Q_{\text{guess}}(\vec{x}|\vec{y},|\psi_S\ra,\{M_{S}^{x_i}\}_i,E)
\end{split}
\end{equation}
\end{proof}

\section*{References}

\providecommand{\noopsort}[1]{}\providecommand{\singleletter}[1]{#1}%

\end{document}